\documentclass{ws-ijmpa}
\usepackage{amsmath}
\usepackage{graphicx}
\def\p{\partial}
\def\s{\sigma}

\def\g{\gamma}

\def\d{\delta}

\def\D{\Delta}

\def\ov{\overline}
\def\ld{\lambda}
\def\Ld{\Lambda}

\def\th{\theta}
\def\e{\eta}

\def\b{\beta}

\def\a{\alpha}
\def\ast{\alpha^*}

\def\pdellx'{\frac{\partial}{\partial x'}}
\def\pdellw'{\frac{\partial}{\partial w'}}

\newcommand{\be}{\begin{equation}}
\newcommand{\ee}{\end{equation}}
\def\bed{\begin{displaymath}}
\def\eed{\end{displaymath}}
\def\bea{\begin{eqnarray}}
\def\eea{\end{eqncrray}}
\def\[{$$}
\def\]{$$}

\begin{document}
\title{Experiments on the Violation of Electromagnetic Gauge Symmetry by Yang-Mills Gravity Using Josephson Effects in Superconductors}
\bigskip
\author{%
 Jong-Ping Hsu\\
  Department of Physics, University of Massachusetts Dartmouth, North Dartmouth, MA  
02747-2300, USA\\
Leonardo Hsu\\
Department of Chemistry and Physics,  Santa Rosa Junior College,\\
Santa Rosa, CA 95401, USA\\}

\maketitle

{\small Yang-Mills gravity is a quantum theory of gravity with translational gauge symmetry that is based on a flat space-time. The universal coupling of all quantum fields to quantum Yang-Mills gravity is based on the replacement of $\p_\mu$ by the translational gauge covariant derivative $(\p_\mu +g\phi_\mu^\nu \p_\nu)$ in the Lagrangians of non-gravitational fields. Near the surface of the Earth, Yang-Mills gravity causes
 the phase gradient $\p_k \th$  to be altered by a factor of $h_1\approx(1- g\phi)\approx 1- 7\times 10^{-10}$.  In addition, the usual gauge-invariant combination of phase gradients and electromagnetic vector potentials ${\bf A}$ in Josephson junctions is modified and is no longer $U_1$ gauge invariant. The voltage across a Josephson junction is thus affected by the presence of the gravitational coupling constant $g$, and is now given by
 $V_{g21}\approx  Q \int_1^2 [- \mbox{\boldmath$ {\bf \nabla}$} A_0 - h_1^{-2}{\p {\bf A}}/{\p t}]\cdot d{\bf s}$.  If one were to compare the voltage across a Josephson junction in a laboratory at rest on Earth with that across a junction in free fall (e.g., in the International Space Station), Yang-Mills gravity predicts a difference on the order of 1 part in $10^{9}$, which should be detectable as the precision of the Josephson junction voltage standard is on the order of a few parts in $10^{10}$.  Measurements of two terms in $V_{g21}$ can test (i) the gravitational effect on the Josephson voltage-phase relation, and (ii) the violation of the $U_1$ gauge symmetry in superconductors by Yan-Mills gravity.



\section{Introduction}

In previous work, my colleagues and I formulated quantum Yang-Mills gravity with spacetime translational ($T_4$) gauge symmetry in inertial frames based on a flat spacetime.  In the geometric-optics limit, the wave equations of quantum particles in Yang-Mills gravity lead to a Hamilton-Jacobi type equation, which involves an effective metric tensor $G_{\mu\nu}(x)$ for classical objects and light rays, which we call the Einstein-Grossmann equation of motion for macroscopic objects. Yang-Mills gravity has been found to be consistent with experiments such as the perihelion shift of Mercury, the deflection of starlight by the sun, redshifts of astronomical objects, and gravitational quadrupole radiation.\cite{1,2,3}  

Yang-Mills gravity has also been quantized in inertial frames and we have calculated the gravitational Feynman-Dyson rules and the S-matrix.\cite{2,4}  This theory has brought gravity back into the arena of gauge field theory and quantum mechanics.  It has also provided a solution to difficulties in physics such as the lack of an operational meaning of local space and time coordinates\cite{5} in Riemannian manifolds, and the incompatibility between `Einstein's principle of general coordinate invariance and all the modern schemes for a quantum mechanical description of nature.'\cite{6} 

\section{Violation of $U_1$ Gauge Symmetry by Universal Gravitational Couplings and the Attractive Force} 

To discuss experiments testing the violation of electromagnetic $U_1$ gauge symmetry by Yang-Mills gravity, we first briefly summarize the key ideas and main equations.   Quantum Yang-Mills gravity is based on an external spacetime translational group ($T_4$) and involves (Lorentz) vector gauge functions $\Ld^\mu(x)$ in flat space-time.\cite{1,2,3}  As a result, the difficulty of quantizing the gravitational field, as discussed by Dyson,\cite{6} disappears.  
The $T_4$ gauge fields are massless spin-2 symmetric tensor fields $\phi_{\mu\nu}$. They are associated with  the $T_4$ group and its generators $p^{\mu}=i \p^{\mu}, \ \hbar=c=1,$ in inertial frames. The $T_4$ gauge covariant derivatives are defined through the following replacement,\cite{1,2,3}
\be
\p_{\mu} \to \p_{\mu} - i g\phi_{\mu\nu}p^{\nu} \equiv J_{\mu\nu}\p^{\nu},  \ \ \ \  \ \ 
J_{\mu\nu}= (\e_{\mu\nu} + g\phi_{\mu\nu}),
\ee
$$
\e_{\mu\nu}=(1,-1,-1,-1),   \ \ \ \ \ \    c=\hbar=1,
$$
in the Lagrangians involving non-gravitational fields such as the electromagnetic fields $A_\mu$ and fermion fields $\psi$.

  As usual, the $T_4$ gauge curvature $C_{\mu\nu\a}$ is derived from the commutator of the gauge covariant derivative $J_{\mu\ld}\p^\ld$, 
\be
[J_{\mu\ld}\p^{\ld}, J_{\nu\s}\p^{\s}]=C_{\mu\nu\a}\p^\a, 
\ee
\be
 C_{\mu\nu\a}=J_{\mu\ld}(\p^\ld J_{\nu\a})-J_{\nu\ld}(\p^\ld J_{\mu\a}).
\ee
To see the effect of the violation of $U_1$ gauge symmetry by gravity on the conservation of electric charge, it suffices to concentrate on the action $\int d^4 x L_{em}$ involving only the electromagnetic field $A_\mu$ and a charged fermion field $\psi$, which are coupled to the gravitational field $\phi_{\mu\nu}$.\cite{1,4} 
The action $S_{\phi\psi}$ in Yang-Mills gravity for the tensor field $\phi_{\mu\nu}$ and a charged fermion field $\psi$ in an inertial frame is quadratic in the gauge curvature $C_{\mu\nu\a}$,
\be
S_{\phi\psi}=\int L_{\phi\psi} d^4 x, \ \ \ \   L_{\phi\psi}= L_{\phi} + L_{em},
\ee
\be   
L_{\phi}= \frac{1}{4g^2}\left (C_{\mu\nu\a}C^{\mu\nu\a}- 2C_{\mu\a}^{ \ \ \  \a}C^{\mu\b}_{ \ \ \  \b} \right),
\ee
\be
L_{em} = - \frac{1}{4}(\D_\mu A_\nu - \D_\nu A_\mu)(\D^\mu A^\nu - \D^\nu A^\mu) 
\ee
$$
+\frac{i}{2}\left[\frac{}{}\overline{\psi} \g^\mu (\Delta_\mu+ieA_\mu) \psi 
 - [(\Delta_\mu-ieA_\mu)\overline{\psi}] \g^\mu  \psi\frac{}{}\right] - m\overline{\psi} \psi, \ \ \ \ \   
$$
$$
\D_\mu A_\nu - \D_\nu A_\mu \equiv F_{\mu\nu} , \ \ \  \D_\mu = (\d^\nu_\mu  + g\phi^\nu_\mu) \p_\nu \equiv J_{\mu}^{\nu}\p_{\nu},   \ \ \ \  e<0,
$$
where we have assumed the replacement in equation (1) for a universal gravitational coupling in Yang-Mills gravity.

The generalized Maxwell's wave equations can be derived from (6).  We have
\be
 \p_{\a}(J^{\a}_{\mu}F^{\mu\nu}) =  e \overline{\psi}\g^{\nu} \psi, \ \ \ \ \ \ \  J_\mu^\a  =\d_\mu^\a + g\phi_\mu^\a.
\ee
The wave equation (7) implies a modified continuity equation for the electric current in the presence of Yang-Mills gravity,\cite{3}
 \be
 \p_\nu J^\nu_{tot}=0, \ \ \ \ \ \ \ \   J^\nu_{tot}= e \overline{\psi}\g^{\nu} \psi - g\p_{\a}(\phi^{\a}_{\mu}F^{\mu\nu}),
 \ee
where we have used the identity $\p_\mu \p_\nu F^{\mu\nu} = 0$. 

Evidently, the usual current $e \overline{\psi}\g^{\nu} \psi$ in the electroweak theory and quantum electrodynamics is no longer exactly conserved in the presence of Yang-Mills gravity, as shown in (8). Instead, a total current, composed of both the usual electromagnetic  current and a new `gravity-em current' [corresponding to the second terms in (8)], is conserved.  The new effect involves a dimensionless earth surface potential $g\phi^{0}_{0} \approx -g\phi^{1}_{1} \approx  -g\phi^{2}_{2}\approx  -g\phi^{3}_{3}\approx Gm/r$,\cite{3} which has a magnitude of roughly $7\times 10^{-10}$ near the surface of the Earth.  Such an effect could be tested experimentally by measuring the voltage across a Josephson junction under different gravitational potentials, as we shall discuss below.

Based on the universal gravitational coupling for all physical particles and antiparticles with the replacement (1) in a Lagrangian, quantum Yang-Mills gravity gives an elegant
explanation as to why the gravitational force is always attractive rather than repulsive.\cite{7,3} 
In quantum Yang-Mills gravity, the properties of the attractive force and the violation of the usual $U_1$ gauge symmetry\cite{3} are embedded in the coupling between the gravitational tensor field $\phi_{\mu\nu}$ and the fermion field $\psi$ of an electron at the quantum level in the Lagrangian (6).  Let us consider the  gravitational tensor field $\phi_{\mu\nu}(x)$ and the electromagnetic potential field $A_\mu(x)$ in the gauge covariant derivative  and its complex conjugate in the fermion Lagrangian (6),
\be
\psi :  \ \   (\D_\mu +ie A_\mu +....) = \p_\mu +g\phi_\mu^\nu \p_\nu +ie A_\mu +.... \ \ \ \
\ee
\be
\ov{\psi}: \ \  (\D_\mu +ie A_\mu +....)^* = \p_\mu +g\phi_\mu^\nu \p_\nu - ie A_\mu +....
\ee
The gauge covariant derivative (9) and its complex conjugate (10) appear in the wave equations of the electron (i.e., particles with charge $e<0$)\footnote{Throughout this paper, we take the constant $e$ to represent a negative value.}  and the positron (i.e., antiparticles with charge $-e$), respectively. 

  The electric force between two charged particles is due to the exchange of a virtual photon.  In quantum electrodynamics, this can be pictured as a Feynman diagram with two vertices connected by a photon propagator.  The key properties of the electric repulsive force, $F_e(e^-,e^-) \propto  (+ie)\times (+ie)= - e^2$,  and the attractive force $F_e(e^-, e^+)\propto (+ie)\times (-ie)=+e^2$  are given by the third terms in (9) and in (10).   Thus, the existence of both attractive and repulsive electric forces is due to the presence of $i$ in the electromagnetic coupling.  
In contrast, the Yang-Mills gravitational attractive force, $F_{YMg}(e^-, e^-)\propto (g) \times (g)= +g^2$ and the attractive force $F_{YMg}(e^-, e^+)\propto (g) \times (g) =+g^2,$ are given by the second terms in (9) and in (10).
   Because the gravitational coupling terms in (9) and (10) do not involve $i$, the gravitational force is always attractive rather than repulsive.
 Finally, the gravitational coupling constant $g$ in (9) and (10) has the dimension of length (in natural units), in contrast to all other coupling constants of fields associated with internal gauge groups, so that $g^2$ is related to Newtonian constant $G$ by $g^2=8\pi G$.\cite{3}    These basic results  appear to indicate  that the space-time translation gauge group of Yang-Mills gravity is just right for gravity,  even though electromagnetic $U_1$ gauge symmetry is violated.

In the macroscopic (i.e., geometric-optics) limit, the wave equations of quantum particles with mass $m$ in Yang-Mills gravity reduces to  a Hamilton-Jacobi type equation, 
\be
G^{\mu\nu}(x)(\p_\mu S)(\p_\nu S)-m^2=0, \ \ \ \  G^{\mu\nu}=\e_{\a\b}J^{\a\mu} J^{\b\nu},
\ee
which is derived in Yang-Mills gravity and is called the Einstein-Grossmann equation  for classical objects.\cite{3}  As a result, the universal gravitational couplings (1) lead to a new viewpoint:  namely that the apparent curvature of macroscopic spacetime appears to be a manifestation of the flat spacetime translational gauge symmetry for the wave equations of quantum particles in the geometric-optics limit.\cite{2,7}  According to quantum Yang-Mills gravity, macroscopic objects move as if they were in a curved spacetime because their equation of motion involves the `effective metric tensor' $G^{\mu\nu}(x)$, which is actually a function of the $T_4$ tensor gauge fields $\phi_{\mu\nu}$ in flat space-time.

 \section{The Josephson Effect Modified by Yang-Mills Gravity}
 The universal gravitational coupling (1) implies that the usual gauge-invariant Dirac and Schr$\ddot{o}$dinger wave equations will be modified by the replacement (1).\cite{3} As usual, we assume that the quantum equation for Cooper-pairs is more or less like the Schr$\ddot{o}$dinger equation, with the difference being that the charge $Q$ will be twice the charge of an electron.\cite{8,9} For purposes of analyzing a Josephson junction, it suffices to consider the Schr$\ddot{o}$dinger wave equation for superconductors with suitable modifications by Yang-Mills gravity:\cite{9}\footnote{How Einstein gravity or general relativity affects the wave equation (12) is unknowable because quantum mechanics and General Relativity (based on a curved space-time) are incompatible.} 
\be
 i h_0\frac{\p \psi}{\p t} = \frac{1}{2M} (-ih_1 \p_k  - QA_k)^2 \psi +Q A_0 \psi,  \ \ \ \ \  
 \ee
where $\p_k=(\p_x,\p_y,\p_z),$ etc. 
 On the surface of the Earth, with $r=R_E$ and mass $m_E$, the non-vanishing components $J^\mu_\mu$ (to the first order approximation) are\cite{3}
\be
J^0_0 =1+g\phi^0_0 \equiv h_0, \ \ \    J^1_1=J^2_2=J^3_3=1- g\phi^0_0\equiv h_1; 
 \ee
$$
 g\phi_0^0 =\frac{Gm_E}{ R_E}\approx  6.95\times 10^{-10},    
 \ \ \ \ \  G=g^2/(8\pi).
 $$ 
 The wave equation (12) violates the $U_1$ gauge symmetry due to the presence of Yang-Mills gravity with $h_0$ and $h_1$.

In the presence of Yang-Mills gravity, the modified charge (probability) current density ${\bf j}_g$ can be derived from (12) and $\p (Q\psi^*\psi)/\p t=-\p_k j_{gk}$.  We have
$$
j_{gk} =  \frac{Q h_1}{2iM h_0} [\psi^* (h_1\p_k \psi - iQA_k \psi )
$$
\be
-  (h_1 \p_k \psi^* + iQA_k \psi^* ) \psi],
\ee
where $k=x,y,z$.  The charge $Q=2e$ and mass $M=2m_e$ in (12) and (14) are used for the Cooper pairs in superconductors.  The relation $Q=2e$ has been confirmed by experiments of flux quantization and the basic flux unit, but $M=2m_e$ has not yet been tested.\cite{8}

To see the effect of Yang-Mills gravity on the supercurrent, let us consider a tunneling potential barrier  $V(x)=V_o$ in the presence of Yang-Mills gravity.  
In order to estimate the effect of Yang-Mills gravity on the DC voltage standard, let us consider the modification of the Josephson voltage-phase relation by gravity, taking into account the tunneling in a basic Josephson junction only.  For this purpose, we consider as usual a static  case involving the vector potential $A_k$ and a constant voltage\cite{9} $V_o$ in the presence of Yang-Mills gravity,
\be
 -\frac{1}{2M} (h_1 \p_k - iQA_k)^2 \psi({\bf r}) = (h_0 E_o - V_o) \psi({\bf r}).
\ee
Using the relations
\be
 \psi({\bf r}) = \psi_d({\bf r}) exp\left[i\frac{Q}{h_1}\int_s^{{\bf r}}{\bf A \cdot ds}\right] ,
\ee
$$
  (h_1 \p_k - iQA_k)^2 \psi({\bf r}) =exp\left[i\frac{Q}{h_1}\int_s^{{\bf r}}{\bf A \cdot ds}\right](h_1   \p_k)^2 \psi_d({\bf r}),
$$
where the inertial point $s$ is arbitrary, the end point of the line integral is ${\bf r}$ and the curl of ${\bf A}$ vanishes.\cite{10}  
  The wave function $\psi_d({\bf r})$ in the insulating region satisfies\cite{9}
\be
 -\frac{ h_1^2}{2M} \p_k^2 \psi_d({\bf r}) = (h_0 E_o - V_o) \psi_d({\bf r}), \ \ \ \ |x| \le a,
\ee
where $V_o > h_0 E_o$, and its solution (for the 1-dimensional case) is
\be
\psi_d(x)=G_1 cosh(x/x_g) + G_2 sinh(x/x_g), \ \ \   
\ee
$$
 x_g=\frac{ h_1}{\sqrt{2M(V_o- h_0 E_o)}}.
$$
 in the insulating region. The boundary conditions at $x=-a$ and $x=a$ for $\psi({\bf r})$ in (16) are
\be
\psi(-a) = \sqrt{n_2} e^{i\th_2}, \ \ \  \psi(+a) =\sqrt{n_1} e^{i\th_1},
\ee
where $n_2$ $(n_1)$ is the constant Cooper pair density at $x=-a$ $(+a)$, etc.  The relations (16), (18) and (19) allow one to solve for the coefficients $G_1$ and $G_2$,
\be
G_1=\frac{\sqrt{n_2} exp[({i\th_2}-iI(-a)] + \sqrt{n_1} exp[{i\th_1}-iI(+a)]}{2 cosh(a/x_g)},  
\ee
$$
G_2= \frac{-\sqrt{n_2} exp[({i\th_2}-iI(-a)] + \sqrt{n_1} exp[{i\th_1}-iI(+a)]}{2 sinh(a/x_g)},
$$
$$
I(-a)\equiv i\frac{Q}{h_1}\int_s^2 {\bf A}\cdot d{\bf s},  \ \ \ \  I(+a)\equiv i\frac{Q}{h_1}\int_s^1 {\bf A}\cdot d{\bf s}, 
$$
 
  In the presence of Yang-Mills gravity, the modified current density $j_{gk}$ in (14) can be written as
\be
j_{gk}=\frac{Q h_1}{2M h_0} Re(-\psi^*[h_1 \p_k +iQA_k]  \psi).
\ee
Since we are interested in estimating the magnitude of the modification to the current-phase relation by gravity, it suffices to obtain the effects on the 1-dimensional relations, $j_{gk}=j_{gx}=j_{g}$ and $ {\bf A}\cdot d{\bf s}=A_x dx$,\cite{9}
$$
j_g=\frac{Q h^2_1}{M x_g h_0} Im(G_1^* G_2).
$$
The modified current-phase relation by Yang-Mills gravity is  
 \be
 j_{g}=j_{gc} sin (\th_{g21}), \ \ \  j_{gc}=\frac{Q h^2_1 \sqrt{n_1 n_2}}{M x_g h_0 sinh(2a/x_g)},
 \ee
 \be
 \th_{g21} =\th_2 -\th_1 - \frac{Q}{h_1} \int_1^2  {\bf A}\cdot d{\bf s},
 \ee
where $j_{gc}$ is the modified maximum supercurrent density, and the integral of  $ {\bf A}\cdot d{\bf s}=A_x dx$ is to be taken across the junction.  
In the presence of Yang-Mills gravity, the rate of change of the modified phase (23) is
\be
h_0 \frac{\p \th_{g21}}{\p t}=h_o \frac{\p \th_{2}}{\p t} - h_0\frac{\p \th_{1}}{\p t} -\frac{Qh_0}{h_1} \int_1^2\frac{\p {\bf A}}{\p t}\cdot d{\bf s}
\ee 
where  $\th_1$ and $\th_2$ are the phases $\th$ in (19) of the two superconductors coupled by the weak link. 
  At the boundary of the electrode,\cite{9} one has
  \be
h_0 \frac{\p \th}{\p t}= -\frac{M}{2Q^2 \rho_o}{ j_g}^2 - Q A_0.  
  \ee
Using the property $(j_{g1})^2=(j_{g2})^2$ in the electrodes, we obtain the relation modified by the Yang-Mills gravitational interaction, 
\be
h_0\frac{\p \th_{g21}}{\p t}= Q\int_1^2\left(-\mbox{\boldmath$ {\bf \nabla}$}A_0 -\frac{h_0}{h_1}\frac{\p {\bf A}}{\p t} \right) \cdot d{\bf s}.
\ee

Thus, in the presence of Yang-Mills gravity, the modified Josephson equation (26) for voltage takes the form
\be
V_{g21}=  h_0 \frac{\p \th_{g21}}{\p t} 
\approx  Q \int_1^2 \left[- \mbox{\boldmath$ {\bf \nabla}$} A_0 - (h_1)^{-2}\frac{\p {\bf A}}{\p t}\right]\cdot d{\bf s}
\ee

In the absence of Yang-Mills gravity, i.e., $h_0=h_1=1$, the modified voltage $V_{g21}$ in (27) reduces to the usual Josephson equation for the voltage $V_{21}$,
\be
V_{21} =  Q  \int_1^2  \left[- \mbox{\boldmath$ {\bf \nabla}$} A_0 - \frac{\p{\bf A}}{\p t}\right]\cdot d{\bf s}=Q \int_1^2 {\bf E}\cdot d{\bf s}.
\ee
We note that (28) is invariant under the usual electromagnetic $U_1$ gauge transformations 
$$
{\bf A}(x)\to {\bf A}'(x) ={\bf A}(x) + \mbox{\boldmath$ {\bf \nabla}$} \Ld(x),
$$
$$
 A_0(x) \to A'_0(x)=A_0(x) - \p \Ld(x)/\p t, 
$$ 
where $x$ denotes any space-time point.  In contrast, (27) is not $U_1$ gauge invariant because $h_0/h_1\approx (h_1)^{-2}$ is not equal to 1 due to the presence of Yang-Mills gravity.

\section{Experiments and Discussions}
The results for voltages in (27) and (28) show the effect of Yang-Mills gravity on the Josephson effect.  The superconductive phenomena involve magnetic fields, which are described by vector potentials.\cite{8}  When the situation is arranged such that the modified voltage (27) is dominated by the vector potential $(h_0/h_1) \p {\bf A}/\p t$,  then Yang-Mills gravity predicts that the measured values of $V_{g12}$ will increase by a factor of $h_0/h_1 \approx (h_1)^{-2}\approx (1+1.4\times 10^{-9})$, as shown in the modified Josephson voltage-phase relation (27). 

  Then, if one were to compare the voltage across a Josephson junction in a laboratory at rest on Earth with that across a junction in free fall (e.g., in the International Space Station or in a plane maneuvering to simulate zero-gravity such as NASA's now-retired ``Vomit Comet"), Yang-Mills gravity predicts a difference on the order of 1 part in $10^{9}$, which should be detectable as the precision in the Josephson junction voltage standard is on the order of a few parts in $10^{10}$.\cite{11,12}

Quantum Yang-Mills gravity with the universal gravitational coupling defined by (1) is essential to understanding why the gravitational force is always attractive\cite{3,7} for all particles and antiparticles. It is formulated on the basis of translational  ($T_4$)  gauge symmetry in a flat space-time, which brings our description of gravity back into the realm of gauge field theory and quantum mechanics. 
 The proposed precision experiment with the Josephson effect can test not only the universality of gauge symmetries with external gauge
group in particle physics, but more importantly the effects of the coupling of the
gravitational and electromagnetic fields, including a violation of electromagnetic
gauge invariance by quantum Yang-Mills gravity.

  The work was supported in part by Jing Shin Research Fund and Prof. Leung Mem. Fund.  
   \bigskip  
 
\bibliographystyle{unsrt}


\end{document}